\documentclass[aps,prl,reprint,longbibliography,superscriptaddress]{revtex4-1}

 \usepackage{amsmath,bm}
 \usepackage{mathrsfs}
 \usepackage{amsfonts}
 \usepackage{graphicx}
 \usepackage{setspace} %leaves captions single space in draft mode
 \usepackage{graphicx}
 \usepackage{epstopdf}
 \usepackage{dcolumn}
 \usepackage{amsmath}
 \usepackage{epsfig}
 \usepackage{indentfirst}
 \usepackage{psfrag}
 \usepackage{subfigure}
 \usepackage{amssymb}
 \usepackage{color}
 \usepackage{units} % rz

 \usepackage{graphicx}% Include figure files
 \usepackage{dcolumn}% Align table columns on decimal point
 \usepackage{bm}% bold math
 \usepackage{natbib}

\usepackage{physics}
\usepackage{dcolumn}% Align table columns on decimal point
\usepackage{bm}% bold mathhttps://www.overleaf.com/project/5af95d0f3775594d14d4a052
\usepackage{natbib}
\usepackage[backref=none,bookmarksnumbered=true,bookmarks=true,bookmarksopen=true,colorlinks=true,
citecolor=blue,linkcolor=blue,anchorcolor=green,urlcolor=blue,unicode=false]{hyperref}

\usepackage{ulem}[normalem] %Whatch out, places underlining in Journal references. Use \normalem before references to deactivate this 

\makeatletter
\newcommand\colorsout[1]{\bgroup \markoverwith{\textcolor{#1}{\rule[0.5ex]{2pt}{0.4pt}}}\ULon}

\makeatother
\begin{document}

\title{Coupled Yu-Shiba-Rusinov states induced by a many-body molecular spin on a superconductor }

\author{Carmen Rubio-Verd\'u }
  \affiliation{CIC nanoGUNE-BRTA, 20018 Donostia-San Sebasti\'an, Spain}
    \affiliation{Department  of  Physics,  Columbia  University,  New  York,  New  York  10027,  United  States}

\author{Javier Zald\'ivar}
  \affiliation{CIC nanoGUNE-BRTA, 20018 Donostia-San Sebasti\'an, Spain}

\author{Rok \v{Z}itko}
 \affiliation{Jo\v{z}ef Stefan Institute, Jamova 39, SI-1000 Ljubljana, Slovenia}
 \affiliation{Faculty of Mathematics and Physics, University of
 Ljubljana, Jadranska 19, SI-1000 Ljubljana, Slovenia}

\author{Jose Ignacio Pascual}
  \affiliation{CIC nanoGUNE-BRTA, 20018 Donostia-San Sebasti\'an, Spain}
\affiliation{Ikerbasque, Basque Foundation for Science, 48013 Bilbao, Spain}

\begin{abstract}
A magnetic impurity on a superconductor induces Yu-Shiba-Rusinov (YSR) bound states, detected by tunneling spectroscopy as long-lived quasiparticle excitations inside the superconducting gap. Coupled YSR states constitute basic elements to engineer artificial superconducting states, but their substrate-mediated interactions are generally weak.     In this paper, we report that intramolecular (Hund’s like) exchange interactions produce coupled YSR states across a molecular platform. We measured YSR spectra along a magnetic iron-porphyrin on Pb(111) and found evidences of two orbital interaction channels, which invert their particle-hole asymmetry across the molecule. Numerical calculations show that the identical YSR asymmetry pattern of the two channels is caused by two spin-hosting orbitals with opposite potential scattering and coupled strongly. Both channels can be similarly excited by tunneling electrons into each orbital, depicting a new scenario for entangled superconducting bound states using molecular platforms. 
\end{abstract}
\date{\today}

\maketitle

A magnetic impurity placed on a superconductor creates a spin-dependent scattering potential that locally distorts the bath of Cooper pairs and creates Yu-Shiba-Rusinov (YSR) bound states \cite{Yu,Shiba,Rusinov}.  YSR states are localized around the impurity and can be  probed by tunneling electrons or holes as quasiparticle excitations, showing in tunneling spectra as pairs of narrow peaks inside the superconducting gap \cite{Ji2008,Heinrich2018}. Interacting YSR states are the basis for engineering novel superconducting states \cite{Morr2003}, with the perspective of manufacturing atomic scale version of the Kitaev chain  \cite{Kitaev,Choy2011,Nadj-Perge2013,Klinovaja2013}.  A crucial element for the resulting many-body state is the nature of the exchange interaction between spins. Most works considered substrate-mediated indirect exchange terms, which are generally weak and rely on substrate's nature. Here we show that the strong direct Hund’s exchange interaction between different orbitals of a magnetic molecule also produce a many body YSR state.  

Atomic and molecular species with high spin can give rise to multiple YSR channels, reflecting the manifold of spin-carrying orbitals that interact with the superconducting substrate. Usually, molecules appear with a single YSR channel \cite{Franke2011,Hatter2015,Hatter2017, Kezilebieke2018,Kezilebieke2019,Brand2018,Malavolti2018,Etzkorn2018} due to their  weak hybridization with the substrate. This single channel may appear split due to  intrinsic molecular spin or vibration excitations \cite{Hatter2015,Hatter2017,Kezilebieke2019, Zitko2011, Golez2012}, or by  interactions with other magnetic species \cite{Kezilebieke2018}. A multichannel picture was reported for some 3$d$ transition metals \cite{Ji2008,Ruby2016,Choi2017}. In these systems, the spatial distribution of each YSR state reflected the orbital shape of the corresponding spin-hosting state, thus behaving as independent channels. Despite the crucial role of Hund’s exchange in determining the magnetic ground state of an impurity, its effect in YSR states remains unknown. 

Here, we report that a magnetic iron porphyrin on the superconductor Pb(111) shows evidence of two  YSR channels interacting  via intramolecular exchange interaction. Scanning tunneling spectroscopy reveals two YSR excitations with identical particle-hole asymmetry distribution, that reverses over the molecular center.  With the support of numerical models, we find that this YSR pattern is caused by two spin-carrying molecular states with opposite potential scattering amplitude and linked through Hund’s intramolecular exchange. The resulting many-body state interacts with the substrate via two coupled YSR channels, both distributed according to the combined shape of two molecular orbitals. In the normal state, this many body spin appears as a narrow Kondo resonance in the spectra, with similar intramolecular distribution than the YSR states. The results presented here motivate the use of spin chains and spin-polarized bands in extended molecular systems as alternative routes to produce interacting YSR states in superconductors.

\begin{figure}[t!]
    \includegraphics[width=0.9\columnwidth]{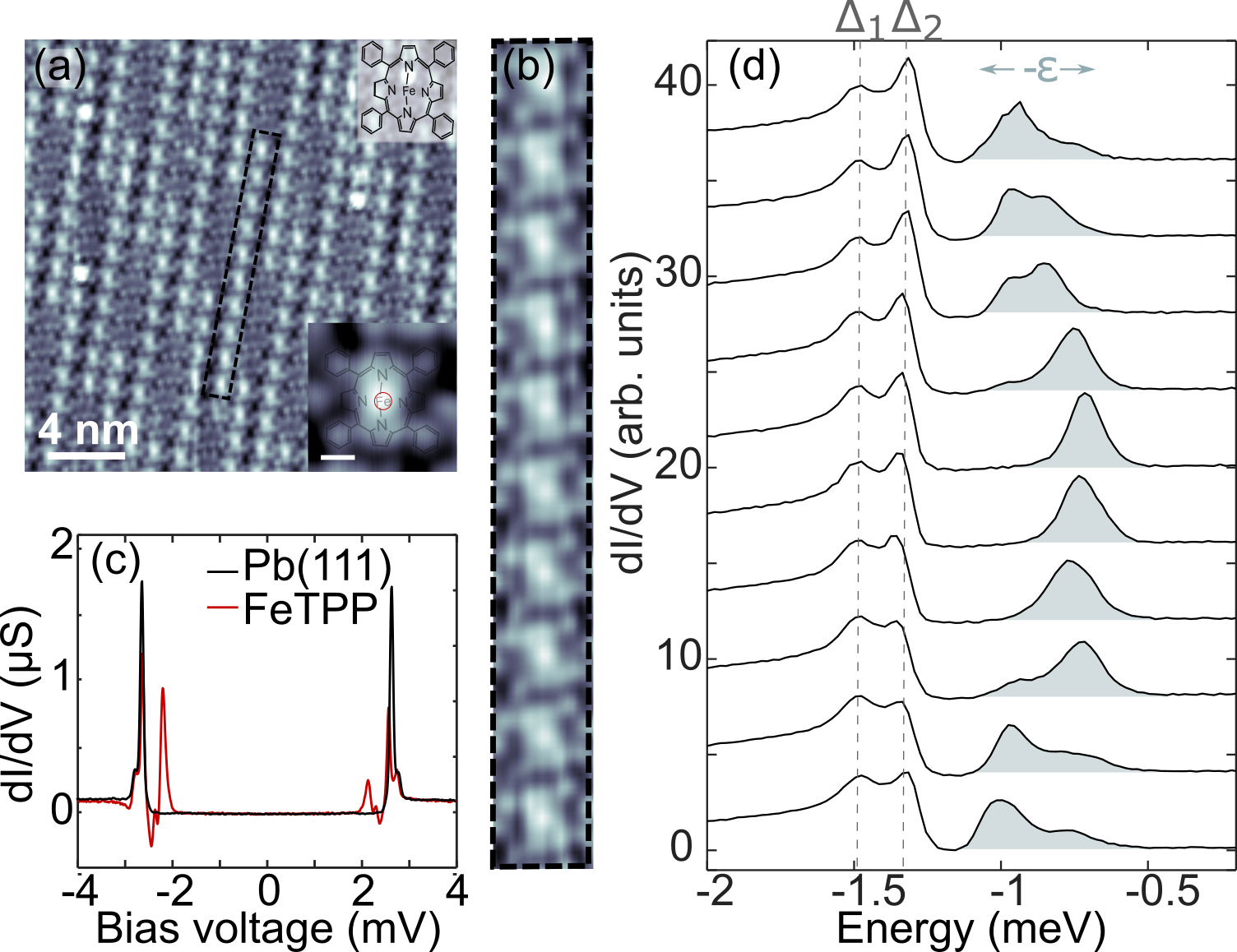}
	\caption{\label{FigTopo} 
	(a) STM topography of a FeTPP island showing the moir\'e pattern. (inset) STM topography of a single bright molecule. The two-fold shape reflects the saddle-shape conformation acquired by the porphine core, with to two pairs of nonequivalent pyrroles \cite{RubioVerdu}. Scalebar represents 4\AA (V$_S = \unit[45]{mV}$, I$=\unit[100]{pA}.$)
	(b) STM topography of a bright molecular segment (V$_S = \unit[45]{mV}$, I$=\unit[100]{pA}.$) 
	(c)  $dI/dV$ spectrum  obtained with a superconducting tip taken over the center of a FeTPP molecule, as indicated with the red circle in a. The bare Pb(111) spectrum is shown in black as a reference (V$_S=\unit[4]{mV}$, I$=\unit[100]{pA}$). 
	(d) $dI/dV$ spectra taken over the center of bright FeTPP molecules along the moir\'e line in b. The plots are obtained by numerically deconvolving the tip's superconducting density of states from $dI/dV$ spectra, as in \cite{Choi2017}.  
	Analysis of STM and spectroscopy data was performed with the WSxM \cite{Wsxm} and SpectraFox \cite{spectrafox} software packages.}
\end{figure}

\begin{figure}[t!]
	\includegraphics[width=0.9\columnwidth]{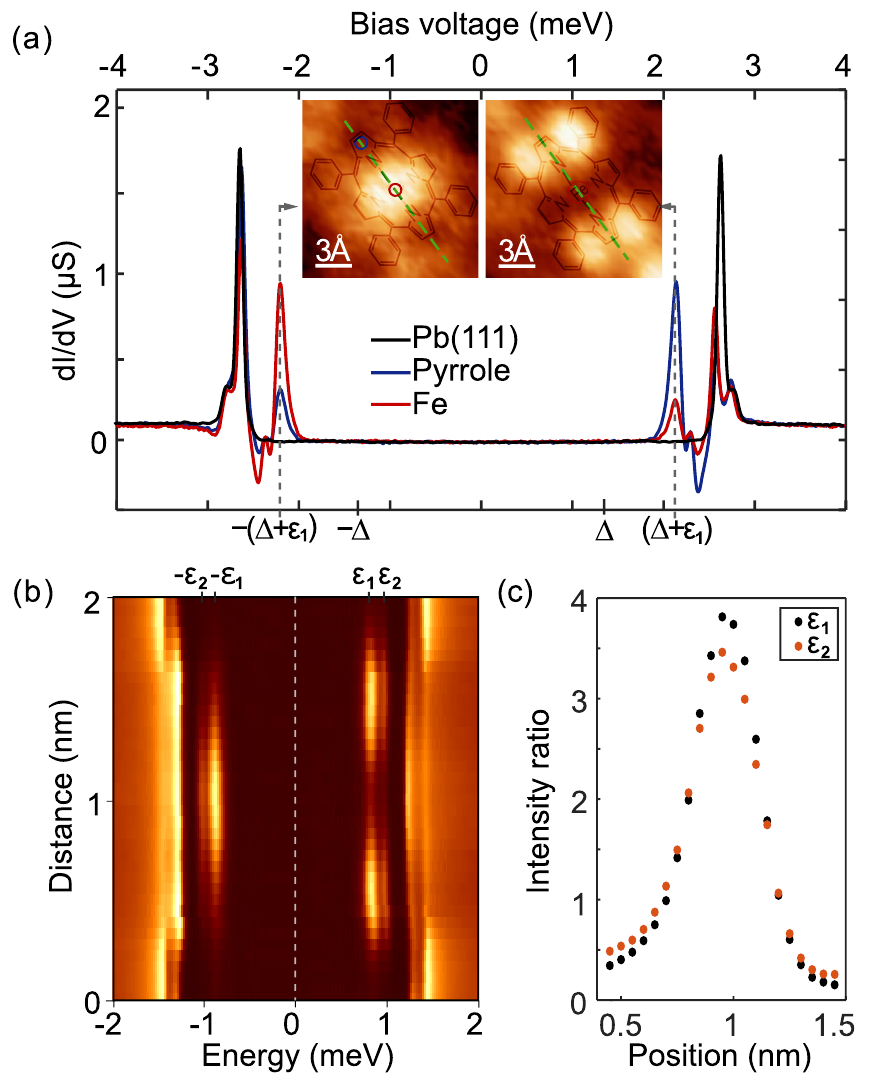}
	\caption{\label{FigShiba} 
	(a) $dI/dV$ spectroscopy on 
	pyrrole (blue) and iron (red) sites of a FeTPP molecule. YSR peaks appear with a pronounced negative differential conductance characteristic of quasi particle tunneling via YSR states \cite{CuevasPRB2020}  Inset: Constant current $dI/dV$  maps at the energy of  hole and electron components of the YSR states, respectively. The locations of the $dI/dV$ curves are shown in circles.  
	(b) Stacking plot of point dI/dV along the FeTPP, numerically deconvolved following ref. \cite{Choi2017}. ($V_S = 	\unit[4]{mV}$, $I = \unit[100]{pA}$). 
(c) The ratio between peak's amplitudes at each polarity ($A_i(V>0)/A_i(V<0$) follows the same trend for both YSR excitations.  
	 }
\end{figure}

Our experiments were performed in a Scanning Tunneling Microscope (STM) at 1.2 K and under UHV conditions (JT-STM by SPECS GmbH). We deposited chlorinated Fe-tetraphenyl porphyrin (FeTPP) molecules on a clean Pb(111) substrate at room temperature, which lose their chlorine ion upon adsorption \cite{Heinrich2013}.  On some metallic surfaces \cite{RubioVerdu,Karan2018}, FeTPP preserves its $S=1$ spin state, albeit  Fe(II) complexes can also adopt a higher integer spin state  \cite{Wang2015,Liu2017,Rolf2018,Rolf2019}. The Pb(111) surface accommodates an incommensurate FeTPP square molecular lattice forming a moir\'e superstructure \cite{Franke2011,Hatter2015}, visible in Fig.~\ref{FigTopo}a as segments of bright and dark molecules.  The dark species show a rather weak hybridization with the substrate, with inelastic spin excitations outside the superconducting gap  \cite{Heinrich2013b,Heinrich2015,RubioVerdu}. In contrast, bright molecules show in-gap states indicative of a stronger hybridization with the substrate, probably caused by a different molecular registry with the Pb(111) atomic lattice \cite{Franke2011}.   

We performed high-resolution $dI/dV$ spectroscopy using a superconducting lead-terminated tip to enhance the energy resolution beyond the thermal limit \cite{Pan1998,Suderow2002,Rodrigo2004,Franke2011} (Fig.~\ref{FigTopo}c). In addition to the characteristic double coherence peaks of the Pb(111) surface at $V_S$ $\approx$ $\pm$ 2.7 mV \cite{Ruby2015b} ($\Delta_1$ and $\Delta_2$ in Fig.~\ref{FigTopo}c), the spectra show two pairs of sub-gap peaks, that we attribute to the excitation of two YSR bound states \cite{Yu,Shiba,Rusinov,Heinrich2018}. The YSR peaks appear at bias voltages spanning from $\pm 2.0$ to $\pm \unit[2.3]{mV}$, depending on the molecule. By deconvolving the tip's superconducting gap  from the spectra (Fig.~\ref{FigTopo}d), we find that the YSR excitation  energies lie in the range $0.5\Delta_1  \lesssim  \epsilon_i  \lesssim  0.7\Delta_1$. 
The two YSR peaks frequently follow the trend shown in Fig. ~\ref{FigTopo}d: their position, separation, and amplitude change from one molecule to another. The energy $\epsilon_i$ is lower in molecular sites around the center of the bright moir\'e segments, and lie closer to the coherence peaks at the ends.  The variations of YSR peaks' position along the lines reflect an increasing overlap of the incommensurate molecular lattice with the lead surface atoms towards the center \cite{Franke2011,Heinrich2013b}: the smaller $\epsilon_i$ values can be attributed to a larger magnetic exchange $J$ between molecular spin and substrate Cooper pairs. This behavior proves that the two peaks correspond to two distinct YSR interaction channels, and excludes  other scenarios providing multiple subgap states, such as a single channel split by intrinsic molecular excitations \cite{Zitko2011,Hatter2015,Pradhan2020,Golez2012}, independent orbital channels with different spatial distribution, as reported for atomic impurities \cite{Ruby2016,Choi2017}, or substrate-mediated indirect exchange with with neighbor molecules \cite{Kezilebieke2018}. 

Both YSR channels display a characteristic asymmetry in the intensity of their particle (p) and hole (h) components. The p-h asymmetry is not homogeneous within the molecule, but exhibits an  intramolecular spatial distribution (see Fig.~\ref{FigShiba}). 
Spectra on pyrrole and Fe sites show the same YSR peaks but the amplitude of their particle
and hole components is reversed (see Fig. \ref{FigShiba}).
Over the pyrrole groups (see Fig.\ref{FigShiba}a), the positive  YSR peaks appear more intense, while  over the Fe ion hole-like excitations ($V_S<0$) are stronger. 
This intriguing  p-h asymmetry pattern is identical for both $\epsilon_1$ and $\epsilon_2$ peaks (Figs. \ref{FigShiba}b,c), and is observed in all the bright molecules of the  moir\'e. 
Such p-h asymmetry is usually interpreted in terms of the hybridization between the low-energy molecular levels and the superconductor, which breaks p-h symmetry in the normal state in the presence of a finite potential scattering amplitude $\mathscr{U}$ \cite{Schrieffer1966,ternes_spin_2015}.   
In terms of the single-impurity Anderson model,  a negative (positive) potential scattering corresponds to  a singly occupied energy level $\epsilon_d$  close (far) to $E_F$ [i.e. $\epsilon_d<U_d/2$ ($\epsilon_d>U_d/2$),  $U_d$ is the Coulomb charging energy] leading to Bogoliubov quasiparticle (BQ) excitations in the superconducting state with larger hole (particle) components
\cite{deGennes1989,Balatsky,Flatte1997,Salkola,Kim2015,Bauer}. 

The asymmetry pattern is consistent with a multi-orbital molecular spin, where the two YSR interaction channels are created by spin-hosting orbitals with  different distribution and energy alignment.   However, the  identical intensity pattern followed by both $\epsilon_1$ and $\epsilon_2$ YSR peaks indicates that they cannot be treated as two independent orbital channels, as in  \cite{Ruby2016,Choi2017}. Instead, we  consider that intramolecular (Hund's-like) exchange  is strong ($J_H \gg \Delta$), forming a robust many-body spin state across multiple orbitals strongly interacting, and coupled with the superconductor via two channels. In this scenario, both  YSR channels can be indistinguishably excited by electrons tunneling though each of the two molecular orbitals.
This case is the superconducting analog of the inelastic excitation pattern of the $S=1$ molecular spin found for Fe(II) complexes on Au(111) \cite{RubioVerdu,Rolf2018}. 

%In the present case, the total spin $S=1$ is fully screened by the superconducting bath into a $S=0$ ground state. There are exactly $N=2$ such sub-gap states, one for each screening channel. Each YSR peak the represents an excitation to an $S^*=1/2$ excited state (partially Kondo screened YSR state with one Bogoliubov quasiparticle bound to the local moment). 

%================= HERE THEORY STARTS

\begin{figure}[t!]
	\includegraphics[width=1\columnwidth]{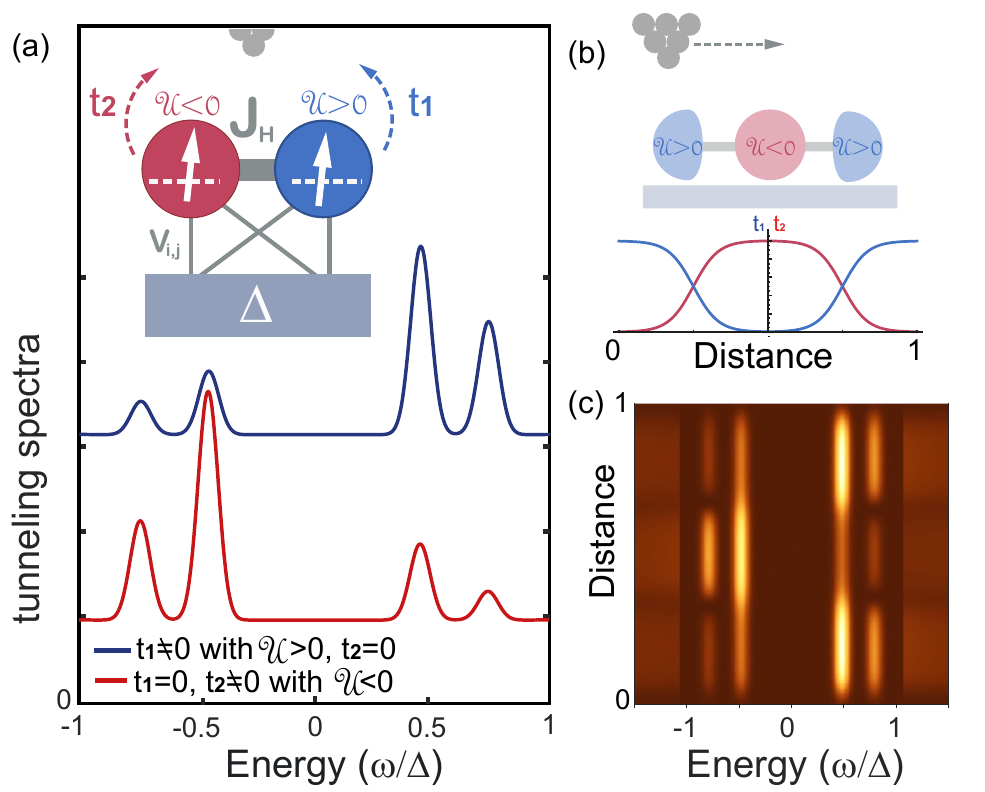}
	\caption{\label{FigTheory} (a) Simulated tunneling spectral functions for two  orbitals with strong Hund's coupling $J_H=0.3$ (in units of the (half)bandwidth) and  with potential scattering  $\mathscr{U}$ of opposite sign (model parameters  $U_{d,1}=U_{d,2}=3$, $\epsilon_{d,1}=-0.25$ and $\epsilon_{d,2}=-2.74$,  the density of states in the band is $\rho=0.5$ and the BCS gap is $\Delta=10^{-3}$). We plot results for the case of both orbitals weakly hybridized with the substrate with matrix elements $V_{1,1} = V_{2,1}$ and $V_{2,2}=-V_{1,2}$, such that $\pi \rho V_{1,1}^2 = 0.05$ and $\pi \rho V_{2,2}^2=0.035$
	The orbitals are coupled with an STM tip via hoping constants $t_1$ and $t_2$.  
	(b) Simulated spatial dependence of the tunneling hopping used in (c) to account for the experimental distribution of p-h asymmetry. (c) Spectral map of YSR excitations with tunneling hopping in (b). }
\end{figure}

To validate such multichannel YSR picture in the presence of local variations of potential scattering,  we simulated low-energy spectra using
numerical renormalization group (NRG) calculations. We considered
two singly-occupied orbitals with opposite potential scattering amplitudes and interacting via
intramolecular exchange coupling to build up a molecular $S=1$ state.
The effective Hamiltonian takes the following form:

\begin{equation}
\begin{split}
H_\mathrm{imp}&=\sum_{n=1,2} \epsilon_{d,n} \sum_\sigma d^\dag_{n,\sigma} d_{n,\sigma}
+ \sum_{n=1,2} U_{d,n} n_{n,\uparrow} n_{n,\downarrow} \\
&- J_H \mathbf{S}_1 \cdot \mathbf{S}_2 +g \mu_B B S_{z,\mathrm{total}}, \\
H_\mathrm{BCS} &= \sum_{i,k\sigma} \epsilon_{k} c^\dag_{i,k\sigma}
c_{i,k\sigma} + \sum_{i,k} \Delta c^\dag_{i,k\uparrow}
c^\dag_{i,k\downarrow} + \text{H.c.}\\
H_\mathrm{hyb} &= \sum_{n,i} \sum_{k\sigma} V_{n,i} d^\dag_{n,\sigma}
c_{i,k\sigma} + \text{H.c.}, \\
H_\mathrm{tunnel} &= \sum_{n,\sigma} t_{n} d^\dag_{n,\sigma}
c_{tip,\sigma} + \text{H.c.},  
\end{split}
\end{equation}

Here $n=1,2$ indexes the molecular orbitals and $i=1,2$
the combinations of substrate electron states that form the different
screening channels; $d_{n,\sigma}$ and $c_{i,k\sigma}$ are the
corresponding operators. The molecule is described by a two-orbital
Anderson impurity model with on-site energies $\epsilon_{d,n}$ and
electron-electron repulsion energies $U_{d,n}$, and an inter-orbital
direct exchange (or Hund's) coupling $J_H$ that aligns the spins
ferromagnetically. The Zeeman term caused by an external magnetic field $B$  is also included. %\mu_B$ teh Bohr magneton 
The operators are $n_{n,\sigma} = d^\dag_{n,\sigma}
d_{n,\sigma}$, $\mathbf{S}_n = (1/2) \sum_{\alpha\beta}
d^\dag_{n,\alpha} \boldsymbol{\sigma}_{\alpha\beta} d_{n,\beta}$ and
$S_{z,\mathrm{total}}=S_{z,1} + S_{z,2}$. 
%We also include an axial magnetic anisotropy term $D$, which is known to be important for this molecular specie \cite{Heinrich2015,RubioVerdu}.  
The substrate is
described by two copies of the BCS Hamiltonian, one for each screening
channel. The hybridization of molecular orbitals and substrate electrons is given by the matrix
elements $V_{n,i}$, which are chosen to obtain YSR
state energies in the experimental range.  

To account for the spatial variations of  $\mathscr{U}$ in the spectral maps, we included a spatial-dependent coupling term with a STM tip through the term $H_\mathrm{tunnel}$ in eq.~1.  
We simulated two orbitals with potential scattering $\mathscr{U}$ of opposite sign, as suggested by the experiments.  
The resulting tunneling spectra [Fig.~\ref{FigTheory}(a)] reproduce the key experimental findings:  the two channels are indistinguishably excited by electron tunneling through either orbital, and both YSR excitations show a clear p-h asymmetry  dictated by the $\mathscr{U}$  of the orbital selected by the STM tip. The spatial variations of p-h asymmetry are readily simulated by inserting a spatial dependence to the tunneling constants $t_n$  [Fig.~\ref{FigTheory}(b)].  The resulting spectral map modulates the YSR amplitudes according to the molecular orbital picked up by the STM tip at every site [Fig.~\ref{FigTheory}(c)], thus reproducing the intramolecular p-h asymmetry observed in the experiments. 

The multi-channel excitation and its response to $\mathscr{U}$ is only obtained in the presence of strong Hund's coupling. We also note that the precise effect of the p-h asymmetry on $\mathscr{U}$ depends on the regime of hybridization amplitude $V_{n,i}$: in the weak regime (free-like spin) the YSR p-h asymmetry reproduces that of the normal state DOS (as in Fig.~\ref{FigTheory}), but this behavior is reversed for channels in the strong-interacting regime (Kondo-screened) \cite{Hatter2015,Farinacci2018}. According to this behavior, the identical p-h asymmetry pattern followed by both channels in the experiment is the result of both lying in the same interaction regime (i.e. we exclude an underscreened-like configuration formed by the combination of a strong channel and a weak one).

%i.e. for $\mathscr{U}<0$ ($\mathscr{U}>0$), the excitation of BQs with larger hole (particle) character leads to more intense YSR peaks at negative (positive) bias. This behaviour is reversed for channels in the strong-interacting regime (Kondo-screened), in which the system undergoes a quantum phase transition to a new ground state in which the YSR level is occupied \cite{Matsuura1977,Balatsky,Franke2011, Heinrich2013b,Farinacci2018}.  In this case, the larger hole (particle) character of the ground state favors now a higher particle (hole) excitation peaks. 

%***********ESTO ES DE VERSION ANTERIOR
%owever, for a given DOS p-h asymmetry, the sign of the YSR asymmetry depends on the strength of the exchange interaction $J$ between YSR impurity and superconductor. In a weak interaction scenario (free spin, $J$ is smaller than $\Delta$)  the YSR p-h asymmetry reproduces that of the normal state DOS, i.e. the excitation of BQs with larger hole (particle) character leads to more intense YSR peaks at negative (positive) bias.  This behaviour is reversed in the strong-interacting regime (Kondo-screened,$J>\Delta$), at which the system undergoes a quantum phase transition to a new ground state in which the YSR level is occupied \cite{Matsuura1977,Balatsky,Franke2011, Heinrich2013b,Farinacci2018}.  In this case, the larger hole (particle) character of the ground state favors now a higher particle (hole) excitation peaks. 

%======================================KONDO

\begin{figure}[b!]
	\includegraphics[width=1\columnwidth]{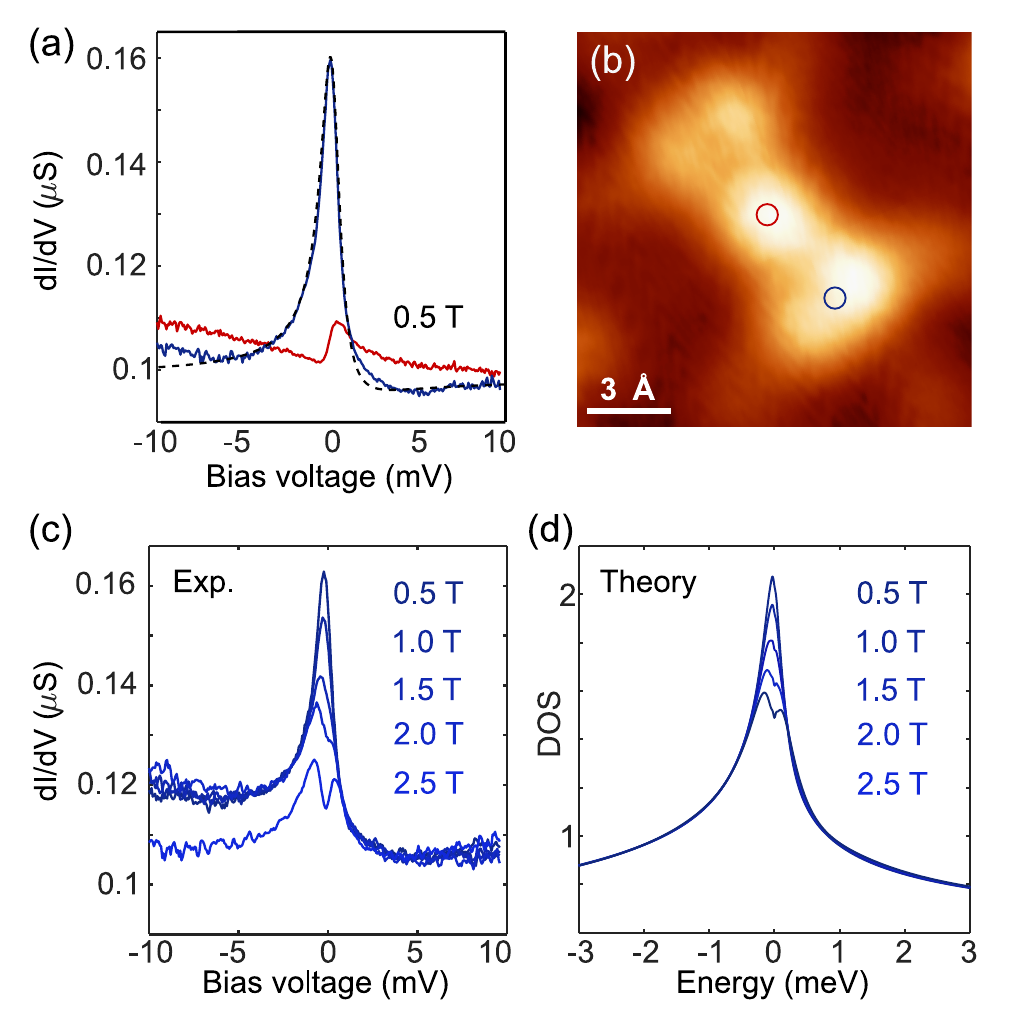}
	\caption{\label{FigKondo} 
	(a) $dI/dV$ spectroscopy of a
	bright FeTPP molecule once superconductivity has been quenched
	on both tip and sample at $B=\unit[0.5]{T}$. Blue and red curves were measured on Fe and pyrrole sites, respectively, as shown in panel b ($V_S =
	\unit[15]{mV}$, $I = \unit[1]{nA}$).  
	(b) Constant-height conductance map obtained at $V_S =
	\unit[100]{\mu V}$ representing the amplitude distribution of the Kondo resonance. (c) $dI/dV$ spectra of FeTPP molecule with increasing out-of-plane magnetic
	field ($V_S = \unit[15]{mV}$, $I = \unit[300]{pA}$). (d) Simulated spectra and splitting with magnetic field for the parameter set used in Fig.\ref{FigTheory}}
\end{figure}

The interaction regime of the two channels can be interrogated in the normal-state. We quenched superconductivity in
both sample and tip by applying an external magnetic field of $B=\unit[0.5] {T}$. In Fig. \ref{FigKondo}a we compare the $dI/dV$ plots  obtained over the Fe and pyrrole sites of the FeTPP molecule, showing zero-bias features compatible with Kondo
physics. On the pyrroles, a sharp zero-bias resonance appears
in the low-energy spectra. On the center of the FeTPP molecule, the
resonance has a strong asymmetry characteristic of Fano-like processes, but with the same line width as over pyrrole sites.

The Kondo resonance exhibits an enhanced sensitivity to the external magnetic field. A sizable splitting is resolved at a magnetic field of $B^* \approx \unit[2]{T}$ (Fig.~\ref{FigKondo}c), which is consistent with a weak-coupling regime, where the Kondo temperature lies below the experimental temperature.
To prove this, we simulated the spectral function in the normal state using  parametrization regimes that reproduce the positions of the experimental YSR peaks in either weak and  strong hybridization regimes. For the later case, the Kondo linewidth is significantly larger than the experimental one and cannot be split by the applied magnetic field. For the weak hybridization  scenario in Fig.~\ref{FigTheory}, however, there is a narrow resonance due to a weak Kondo interaction regime, thermally broadened \cite{Zhang2013e,Hatter2017}, that splits with $B$, thus reproducing the experimental finding in the normal state (Fig.~\ref{FigKondo}d). 

\begin{figure}[t!]
	\includegraphics[width=1\columnwidth]{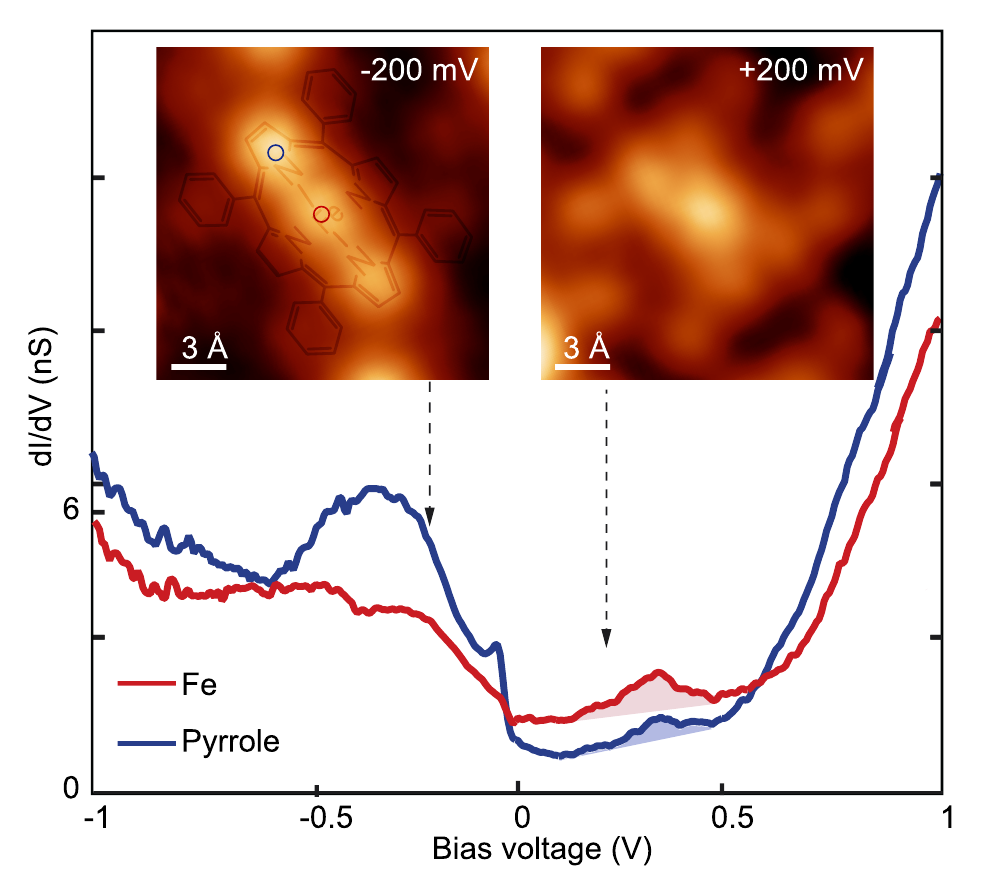}
	\caption{\label{Figorbital} 
	 Wide-range $dI/dV$ spectra of a bright molecule measured on the locations showed in the inset ($V_S = \unit[1.5]{V}$, $I = \unit[3]{nA}$, $V_{\mathrm{rms}} = \unit[2]{mV}$). Inset:  Constant height conductance maps measured at $V_S = \unit[-200]{mV}$ and $V_S=\unit[+200]{mV}$, respectively ($I = \unit[200]{pA}$).  }
\end{figure}

The distribution of the Kondo resonance reveals the shape of the many-body orbital hosting the molecular spin \cite{Li19,Moro-Lagares2019,Liarxiv19}. A spatial map of the Kondo amplitude  (dI/dV at V$\sim$0, Fig.~\ref{FigKondo}b) reproduces an elongated shape composed of a round protrusion over the Fe site and  two-fold symmetric lobes over the bright pyrrole groups. 
This image  resembles the YSR amplitude maps in the superconducting state (inset in Fig.~\ref{FigShiba}(a)), where particle and hole components are localized on pyrrole or Fe sites, respectively. 

This extended Kondo state can be related to frontier molecular states, pictured in Fig.~\ref{Figorbital}. The two YSR channels arise from a finite hybridization of two molecular orbitals with metal states. A molecular resonance at $\unit[400]{meV}$ above E$_F$ appears extended over the pyrrole groups with a shape similar to  Kondo and YSR maps. From its alignment and shape, we conclude that this corresponds to a spin-hosting molecular orbital with $\mathscr{U}>0$, thus responsible for larger YSR particle components over the molecular sides. Below E$_F$ the spectra is dominated by a manifold of molecular states, with strong weight over the organic ligand and over the Fe ion. Here, the most probable scenario is that the Fe ion hosts a second spin-carrying state (as in  \cite{RubioVerdu}) contributing to the total molecular spin. Given its alignment  below E$_F$, this state has opposite potential scattering ($\mathscr{U}<0$) and leads to inverted particle-hole asymmetry over the center.

In summary, the scenario depicted here provides evidence for a novel quantum regime for superconducting bound states induced by a magnetic molecule.  We have shown that the many-body molecular spin of the Fe(II) complex FeTPP interacts with Pb(111) via two Yu-Shiba-Rusinov channels, caused by two frontier molecular orbitals strongly coupled by intramolecular exchange. As a consequence, both channels are indistinguishably excited by tunneling events through each spin-hosting orbital and, therefore, appear with the spatial distribution of the combined spin-hosting molecular states.  These results show that entanglement of YSR states can proceed via intramolecular exchange, as alternative to the weaker substrate mediated exchange interactions. This motivates further studies using spin chains in polymers, or graphene ribbons as possible platforms to produce extended YSR bands. 

% In summary, our results demonstrate the existence of a multichannel Yu-Shiba-Rusinov picture  describing the interaction of the  molecular spin of the  Fe(II) complex FeTPP with the superconducting Pb(111). The particle-hole asymmetry of both pairs of sub-gap Yu-Shiba-Rusinov excitations describes the same intramolecular distribution. With the support of NRG calculations we find that FeTPP molecules have a robust many-body spin interacting with the substrate through two weakly screening channels. In this multichannel configuration  both YSR channels are excited similarly by electron tunneling through each of the molecular orbitals, while their particle-hole symmetry is governed by the tunneling pathway selected by the STM tip.  The scenario depicted here provides evidences for novel quantum regimes for spins on superconductors, where intra-molecular Hund's exchange  govern the magnetic behaviour even in a multichannel hybridization.  

\begin{acknowledgments}
We thank La\"etitia Farinacci, Felix von Oppen, and Katharina Franke for discussions and for sharing with us their work before submission (arXiv:2007.12092). 
J.Z. and N.P. acknowledges financial support from Spanish Agencia Estatal de Investigaci\'on and the European Regional Development Fund (ERDF) (PID2019-107338RB-C1) and MAT2016-78293-C1 and the Maria de Maeztu Units of Excellence Programme MDM-2016-0618), and from the European Union (Horizon 2020 FET-Open project SPRING Grant. no. 863098). C. R. V. acknowledges funding from the European Union’s Horizon 2020 research and innovation programme under the Marie Sk\l odowska-Curie grant agreement No 844271.
R.~\v{Z}. is supported by Slovenian Research Agency (ARRS) under Program P1-0044.
\end{acknowledgments}

%\section*{References}
%\setstretch{1.0}
\bibliographystyle{apsrev4-1} 

%\bibliography{Porphyrin}

\end{document}